\documentclass{ws-rv9x6}
\usepackage{caption}
\usepackage{emptypage}
\usepackage{subfigure}   
\usepackage{ws-rv-thm}   
\usepackage{lscape}

\usepackage{ws-rv-van}   
\makeindex

\begin{document}

\chapter[]{Fully analog memristive circuits for optimization tasks:\\ a comparison}\label{ra_ch1}

\author[F. C. Sheldon,  C. Coffrin, F. Caravelli]{F. C. Sheldon$\ ^{\dagger,\#}$, F. Caravelli$\ ^\dagger$, C. Coffrin$\ ^*$ }

\address{
$\ ^\dagger$T-Division (T4), $\ ^\#$Center for Nonlinear Studies and $\ ^*$A-Division (A1)\\
Los Alamos National Laboratory, Los Alamos, New Mexico 87545, USA
}

\begin{abstract}
We introduce a Lyapunov function for the dynamics of memristive circuits, and compare the effectiveness of memristors in minimizing the function to widely used optimization software. We study in particular three classes of problems which can be directly embedded in a circuit topology, and show that memristors effectively attempt at (quickly) extremizing these functionals.
\end{abstract}
\body

\section{Introduction}

As the challenges of scaling traditional transistor-based computational hardware continue to intensify, ``Moore’s Law,” governing the exponential increase of transistor density, is coming to an end. While the first computers were analog \cite{reviewCarCar}, in the past decades digital computing has made incredible progress and our laptops are now more powerful than the supercomputers just 30 years ago.
On the other hand, there remain hard computational problems that still challenge computer scientists and modern digital computers; in particular many optimization problems. Recently, interest has grown in embedding algorithms directly in analog hardware in the hope that the corresponding hardware speedup could yield a useful specialized processor. In this chapter we focus on the application of analog nanoscale electronic devices with memory,
more specifically memristors. Proposals for specialized co-processors formed of memristors show extreme breadth and versatility in computing applications \cite{Chua76,Yang2013,Chua2014,Strukov2008,reviewCarCar,reviewGorm,diventratravrev,TraversaPol}, ranging from optimization to artificial neural networks.  Here we focus on understanding how the native dynamics of memristive circuits encode features of optimization problems.

Memristors are two-terminal devices that display pinched (at the origin) hysteretic behavior in their voltage-current diagram. Physical memristors \cite{Strukov2008,Yang2013,reviewCarCar} have rather non-trivial voltage-current curves, but many core features are captured by a simple description which we adopt in this paper. In this model, the state of the resistance varies between two limiting values and can be described by a  parameter $w$ which depends on the previous history of the device dynamics and thus may be interpreted as a memory.  We will refer to $w$ as the \emph{internal memory parameter}. In spirit, memristors have the essential property that the underlying dynamics are the result of competition between resistance reinforcement, caused by the flow of currents through the device, and a thermodynamically driven decay \cite{Sheldon2017,Adamatzky2012}. Recent advancements show that there is a deep connection between the asymptotic memory states of the circuits and the solutions of combinatorial optimization, and the ground states of the Ising model and spin glasses.  Additionally, memristors offer a possible substrate to construct neuromorphic chips, e.g. electronic components that behave similarly to human neuronal cells. Central to all of these applications is that memristors, as we  show in this paper, can perform computation without requiring CMOS, thus in a fully analog fashion. As a result, circuits of memristors have been proposed as a potential basis for the next generation of passive and low-energy computational architectures.

Interest in specialized analog co-processors for solving optimization problems has generated a host of possible approaches. While some of these problems can be in principle be tackled using quantum computers \cite{Carletonio,QC}, it is unlikely that these will be available for mass distribution. One of the proposed alternative paradigms is in-memory computation \cite{DiVentra2013b}: removing the separation between memory and computing typical of the von Neumann architecture.  In this approach specialized circuits are designed to utilize active components in concert with memristors to obtain the solution of a specific problem \cite{maxsat,Traversa2015a}.  In this work, we consider a more fundamental question: do the dynamics of circuits of memristors encode optimization problems natively? 
 Understanding their asymptotic behavior requires characterizing the interplay between nonlinear dynamics, interactions and constraints and as a result the dynamics of memristor networks is still an area of active research, despite the the fact that the theory behind a single device was introduced over half a century ago \cite{chua71,Chua76}.

With this purpose in mind, in this paper we study a specific optimization problem in the context of fully analog memristive circuits, e.g. circuits composed only of memristors. For these circuits we can take advantage of an exact evolution equation for the internal memory parameters, which will serve as our case study. For these equations, we derive a novel Lyapunov function (and which solves some of the problems of a Lyapunov function provided in the literature). Being the Lyapunov function being minimized by the memristive network, we compare the results of the minimization to state of the art optimization software.

\section{Dynamical equation for memristor circuits}

\subsection{Single memristor and Lyapunov function}

For the case of titanium dioxide devices, a rather simple toy model for the evolution of the resistance is the following:
$$ R(w)=R_{on} (1-w) +w R_{off}\equiv R_{on}(1+\xi w),$$
\begin{equation}
     \frac{d}{d t} w(t)=\alpha w(t)- \frac{R_{on}}{\beta} i(t),
     \label{eq:memr1}
\end{equation}
initially studied for $\alpha=0$, and where $0\leq w\leq 1$, $\xi=\frac{R_{off}-R_{on}}{R_{on}}$; in the equation above $i(t)$ is the current flowing in the device at time $t$. Physically, $w$ can be interpreted as the level of internal doping of the device, but this is a crude description. The constants $\alpha,\beta$ and $\xi$ control the decay and reinforcement time scales and the degree of nonlinearity in the equation respectively, and can be measured experimentally. While $\xi$ is adimensional and depends only on the resistance boundaries, $\alpha$ has the dimension of an inverse time, while $\beta$ has the dimension of time divided by voltage. Aside from applications to memory devices, there is interest in these components also because memristors can serve as memory for neuromorphic computing devices \cite{volc}. 

We first demonstrate that this equation possesses a Lyapunov function that governs it's asymptotic behavior. In order to understand the Lyapunov function of the full network, we begin with the case of a single memristor driven by a voltage generator $V(t)$. From the equations above, we have 
\begin{equation}
     \frac{d}{d t} w(t)=\alpha w(t)- \frac{R_{on}}{\beta} \frac{V(t)}{R_{on} \big(1+\xi w(t)\big)},
\end{equation}
from which we obtain
\begin{eqnarray}
    \big(1+\xi w(t)\big) \frac{d}{d t} w(t)&=&\alpha\big(1+\xi w(t)\big)w(t)- \frac{1}{\beta} V(t) \nonumber \\
    &=&\alpha \Big( w(t)+ \xi w(t)^2- \frac{V(t)}{\alpha \beta}  \Big) .
\end{eqnarray}
Let us define now 
\begin{eqnarray}
    L(w)=a\ w(t)^2+b\ w(t)^3+c\ w(t) V(t).
\end{eqnarray}
We have
\begin{eqnarray}
    \frac{d}{dt} L(w)=\Big(2 a\ w(t)+3 b\ w(t)^2 +c\  V(t)\Big)\frac{dw}{dt}+c w(t) \frac{d V}{dt}.
\end{eqnarray}
Now assume that $V(t)= V_0$. If we choose
\begin{eqnarray}
    a&=&-\frac{1}{2},\ \ \ \ \ \ b= -\frac{1}{3} \xi ,\ \ \ \ \ \ c= \frac{1}{\alpha \beta}
\end{eqnarray}
Thus
\begin{eqnarray}
\frac{d}{dt} L(w)&=&    \Big(- w(t)-\xi w(t)^2 +\frac{1}{\alpha \beta}\  V_0\Big)\frac{dw}{dt} \nonumber \\
&=&-\alpha \big(\frac{dw}{dt}\big)^2.
\end{eqnarray}
Thus if $\alpha>0$
\begin{eqnarray}
    \frac{d L}{dt}\leq 0\text{   if   } \frac{dw}{dt}\neq 0,
\end{eqnarray}
with
\begin{eqnarray}
    L(w)=\frac{V_0}{\alpha \beta} w(t) -\frac{1}{2} w(t)^2 -\frac{1}{3} \xi w(t)^3.
\end{eqnarray}
Now, for $\alpha=0$ the solution is of the form $w(t)=\frac{\sqrt{1+q V_0 t}-1}{c}$ and thus $\frac{d}{dt}w=0$ can be only satisfied only for $w=1$ or $w=0$.
For $\alpha\neq 0$ there is no explicit analytical solution but it can be expressed in the form
\begin{eqnarray}
    s&=&\frac{V_0}{\beta} \nonumber \\
    q(t)&=&c_0-t \nonumber \\
    f(t) &=&\frac{\log \left(\alpha  \xi  q(t)^2+\alpha  q(t)+s\right)}{2 \alpha }+\frac{\tan
   ^{-1}\left(\frac{\sqrt{\alpha } (2 \xi  q(t)+1)}{\sqrt{4 \xi  s-\alpha
   }}\right)}{\sqrt{\alpha } \sqrt{4 \xi  s-\alpha }} \nonumber \\
   w(t)&=&f^{-1}(t) \nonumber \\
    1\geq w(t)&\geq& 0,
\end{eqnarray}
whose analysis goes beyond the scope of this paper. 

However, a way to see that the system must eventually reach one of the boundary points $w=\{1,0\}$, is the fact that there is  fixed point for the dynamics, which is defined by the equation
\begin{eqnarray}
     w^* (1+\xi w^*)=\frac{V_0}{\alpha \beta}.
\end{eqnarray}
However, the analysis of the stability of the fixed point reveals that this is an \textit{unstable} fixed point.  From this fact we can intuitively understand that if $w(0)>w^*$, necessarily we have $w(\infty)=1$, and while if $w(0)<w^*$ we obtain $w(\infty)=0$. A similar analysis applies to the case of a network of connected memristors, as we will see shortly.

Given the fact that $w(\infty)\in\{1,0\}$, we have $w^n(\infty) = w(\infty)$ and we can simplify the asymptotic form of the Lyapunov function to
\begin{eqnarray}
    L(w_{\infty})&=&\frac{V_0}{\alpha \beta} w_{\infty} -\frac{1}{2} w_{\infty}^2 -\frac{1}{3} \xi w _{\infty}^3 \nonumber\\
    &=&(\frac{V_0}{\alpha \beta} -\frac{1}{2} -\frac{1}{3} \xi) w_{\infty}.\nonumber
\end{eqnarray}
This function has asymptotic values
\begin{eqnarray}
    &=&\{\frac{V_0}{\alpha \beta} -\frac{1}{2} -\frac{1}{3} \xi,0\}\nonumber \\
    &=& \{w^* (1+\xi w^*)-\frac{1}{2}-\frac{1}{3}\xi,0\}.
\end{eqnarray}

The dynamics of a memristor are thus connected to an optimization problem of the form,
\begin{eqnarray}
    L^*=\text{min}\{\frac{V_0}{\alpha \beta} -\frac{1}{2} -\frac{1}{3} \xi,0\},\text{  or    } 
    L^*=\text{min}\{\frac{V_0}{\alpha \beta}  -\frac{1}{3} \xi,\frac{1}{2}\}, 
    \label{eq:minlyap}
\end{eqnarray}
however we have no guarantee that the dynamics will ``pick" the correct minimum of the Lyapunov function and from our analysis above, we see that this should be depend on the initial conditions.
It is easy to perform simulations of the system above. For instance, we find that for $\alpha=0.1$, $\beta=\xi=10$, and $V=0.92$, the system ends in the real minimum of the asymptotic function 70\% of the time, yet still the system can have a macroscopic portion of asymptotic states not in the minimum of the Lyapunov ``energy". This fact shows that while the Lyapunov function is being minimized along the dynamics of the memristors, the system can effectively be trapped in local minima. This is why in this paper we focus on the minimization of a continuous Lyapunov function for which we can compare the observed asymptotic states from the memristor dynamics to minima obtained via state of the art optimization software.


\subsection{Circuits}

We now with to extend the analysis we did for a single memristor to a circuit.  We consider a graph in which each edge contains a memristor and voltage generator in series.  The state of the internal memory parameters is thus a vector $\vec{w}$ in which each entry corresponds to an edge and each are driven by voltage generators $\vec{s}(t)$.  Memristors in the graph will now interact due to shared currents at the nodes/electrical junctions of the graph.

The extension of eqn. (\ref{eq:memr1}) to a circuit can be done, and is given by
\begin{equation}
\frac{d}{dt}\vec{w}(t)=\alpha\vec{w}(t)-\frac{1}{\beta} \left(I+\xi \Omega W(t)\right)^{-1} \Omega \vec s(t),
\label{eq:diffeq}
\end{equation}
with the constraints $0 \leq w_i\leq 1$ and where we use the convention that $W(t) = diag(\vec{w}(t))$ is a diagonal matrix containing the internal memory parameters \cite{Caravelli2017,Caravelli2017b}. The projection operator  $\Omega_{ij}$ contains the information about the topology of the graph and can be thought of as picking out configurations consistent with Kirchoff's voltage law. As we will discuss shortly, components of $\Omega_{ij}$ may also be considered as the interaction strength between memristors in the graph.
We note that because $\Omega$ is a projection operator, $\Omega=\Omega^2$ we can always write $\vec s=\Omega \vec s+(I-\Omega) \vec s$, it is straightforward to show that we can add to $\vec s$ any vector $\tilde s=(I-\Omega) \vec k$, which will not affect the dynamics. This form of freedom arises from the Kirchhoff constraints from which the differential equation has been derived. 

The set of coupled differential equations above incorporate all dynamical and topological constraints of the circuit exactly \cite{Caravelli2017,Caravelli2017b}. Kirchoff's Laws manifest themselves via the projection operator $\Omega$ which intervenes in the dynamics. Such projector operator also emerges for purely resistive circuits with edges of the graph containing voltage generators $S_i$ in series to resistors $r_i$. For the case of constant resistance $r_i = r$, the equilibrium currents can be written in a vectorial form as 
    \begin{equation}
        \vec i(t)=- \frac{1}{r}\Omega \vec S(t),
    \end{equation}
where $\Omega=A^t (A  A^t)^{-1} A $ is a non-orthogonal projector on the cycle space of the graph.  The matrix $A$ has the dimension $Cycles\times Edges$ of the graph (each row designates a fundamental cycle of the graph), and thus $\Omega$ has the correct dimension (e.g. the number of memristors).\cite{Caravelli2017,zegarac}

We can also generalize equation (\ref{eq:diffeq}) to various forms of driving including current generators in parallel with memristors or current/voltage generators driving the nodes of the circuit.  We cast these in a general form using a generic source vector $\vec{x}$ as,
\begin{equation}
\frac{d}{dt}\vec{w}(t)=\alpha\vec{w}(t)-\frac{1}{\beta} \left(I+\xi \Omega_A W(t)\right)^{-1} \vec x,
\label{eq:diffeqgen}
\end{equation}
where we have
\begin{eqnarray}
    \vec x=\begin{cases}
    \Omega_A \vec s & \text{Voltage sources in series}\\
    A(A^T A)^{-1} \vec s_{ext}  & \text{Voltage sources at nodes}\\
    \Omega_B \vec j & \text{Current sources in parallel}  \\
    B^T(B B^T)^{-1} \vec j_{ext} & \text{Current sources at nodes}.
    \end{cases}\nonumber
\end{eqnarray}

The purpose of this chapter is to further understanding of the asymptotic dynamics of a circuit of memristors, and specifically the statistics of the resistive states. An analysis of the asymptotic states can be done via Lyapunov functions as we did for the case of a single memristor. After a first attempt at deriving a Lyapunov function \cite{Caravelli2019Ent}, plagued by constraints on the external fields, here we provide a novel yet similar Lyapunov function free of these requirements. From the point of view of optimization with analog dynamical systems, different Lyapunov functions provide different ways of embedding a computational problem in a physical system.

We follow the same prescription as the single memristor case, but where the interaction matrix is a projection operator on the cycle basis of the circuit.

\subsection{Lyapunov function for memristor circuits}
   We begin with the equations of motion,
\begin{equation}
    (I + \xi \Omega W)\dot{\vec{w}} = \alpha \vec{w} + \alpha \xi \Omega W \vec{w} - \frac{1}{\beta} \vec{x}
\end{equation}
where we have multiplied by $(I + \xi\Omega W)$. Consider

\begin{equation}
    L = -\frac{\alpha}{3} \vec{w}^T W \vec{w} - \frac{\alpha\xi}{4} \vec{w}^T W\Omega W \vec{w}
    +\frac{1}{2\beta} \vec{w}^T W\vec{x}.
    \label{newlyap}
\end{equation}
In this case, we have
\begin{align}
    \frac{dL}{dt} &= \dot{\vec{w}}^T\left(-\alpha W\vec{w} - \alpha\xi W\Omega W\vec{w} + \frac{1}{\beta} W\vec{x} \right)\nonumber \\
    {} &= -\dot{\vec{w}}^T (W + \xi W\Omega W)\dot{\vec{w}}\nonumber\\
    &= -\dot{\vec{w} }^T \sqrt{W}(I + \xi \sqrt{W}\Omega \sqrt{W})\sqrt{W}\dot{\vec{w}} \nonumber\\
    &= -||\sqrt{W}\dot {\vec{w}}||^2_{ (I + \xi \sqrt{W}\Omega \sqrt{W})} \label{eq:lyapderiv}
\end{align}
and we have that $\frac{dL}{dt} \le 0$ as $(I + \xi\sqrt{W}\Omega\sqrt{W})$ is positive definite. We thus have that $L$ in equation (\ref{eq:newlyap}) is a Lyapunov function for a circuit of memristors.

An asymptotic form can be obtained by replacing $w_i^k=w_i$ for integer $k$, as asymptotically one has $w_i=\{1,0\}$.
Thus, the asymptotic Lyapunov function form is given by
\begin{equation}
    L(\vec w)  =  - \frac{\alpha\xi}{4} \vec{w}^T \Omega  \vec{w}
     +\vec{w}^T\left(\frac{1}{2\beta}\vec{x} - \frac{\alpha}{3}\right)\nonumber
\end{equation}
which is a form familiar from physics in the context of spin systems.  We can re-express this in terms of spin variables $\sigma_i = 2w_i - 1$ and with a few simplifications as
\begin{equation}
    \tilde L=\frac{8L(\vec \sigma)}{\alpha}=\vec \sigma\cdot (\frac{2\vec{x}}{\alpha \beta} -\frac{4}{3}  \vec 1- \xi   \Omega  \vec 1) - \frac{\xi}{2} \vec{\sigma}\ \tilde \Omega \  \vec{\sigma}
\end{equation}
where $\tilde \Omega$ has only the off-diagonal terms of $\Omega$.
The structure of the Lyapunov function above is very similar to the one described before, but only contains the spectral condition $I+\xi \sqrt{W}\Omega \sqrt{W}\geq 0$, which is natural. We can thus identify an effective local field $\vec{h} = \frac{2\vec{x}}{\alpha \beta} -\frac{4}{3}  \vec 1- \xi   \Omega  \vec 1$ and interactions between memristors given by $\tilde \Omega$.

\begin{figure}
    \centering
    \includegraphics[scale=0.3]{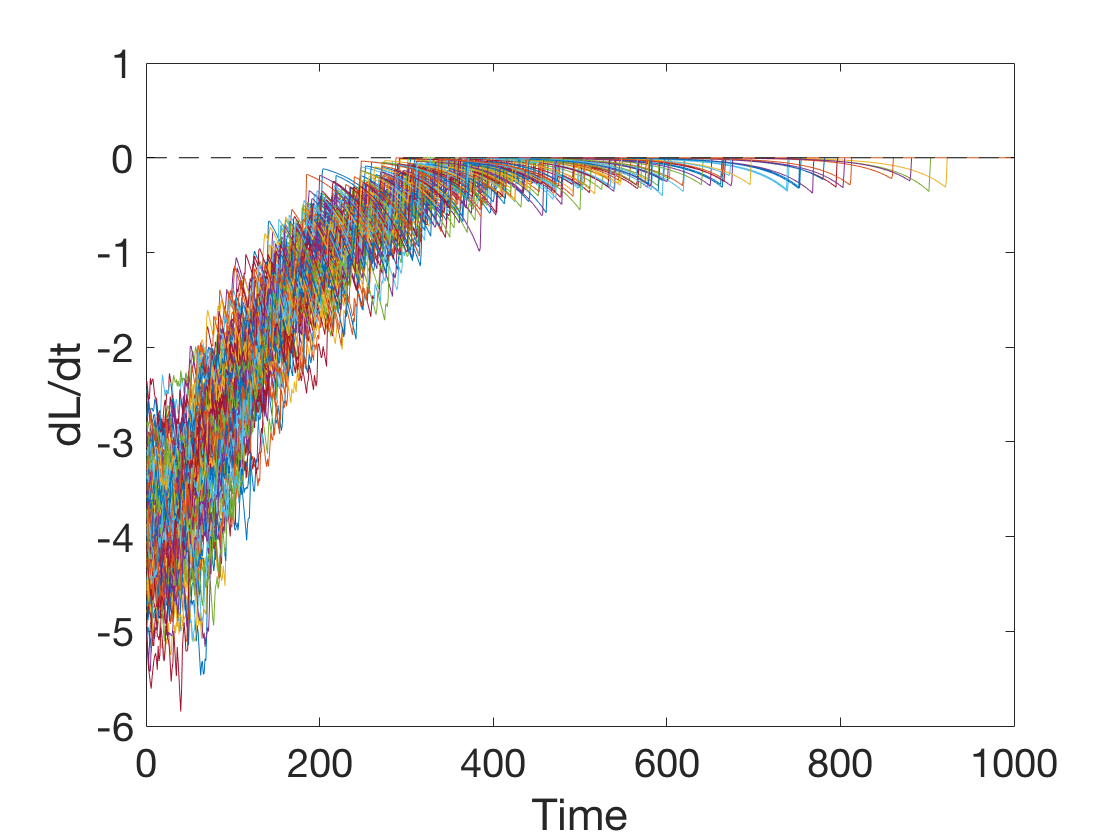}
    \caption{Derivative of the Lyapunov function of eqn. (\ref{eq:newlyap}) for a 100 random initial conditions and instances $(\Omega, \vec h)$. We see that the derivative is always negative, and thus $L$ is decreasing.}
    \label{fig:randominst}
\end{figure}

A notable omission from the Lyapunov function argument above is the presence of boundaries on the internal memory parameters $w_i$.  As individual memristors reach their boundaries and their dynamics halted, components of the derivative in equation (\ref{eq:lyapderiv}) go to 0.  As a test of the fact that the Lyapunov function above works when including boundary effects,
in Fig. \ref{fig:randominst} we plot $\frac{dL}{dt}$ evaluated numerically for 100 instances $( \Omega, \vec h)$, in which $\Omega$ was obtained from random circuits and $\vec h$ is a gaussian-distributed vector. 

We now wish to show that the Lyapunov function converges asymptotically only on the boundary of the set $[0,1]^N$, which is what one observes numerically.

\subsection{Number of fixed points and stability}
As for the case of the one dimensional model, the fixed points of the dynamics are important in order to understand the stability of the system. In the previous section we have assumed that our Lyapunov function can be replaced with an asymptotic form which is on the binary set $w_i=\{0,1\}$. We wish to show this feature in this section.

The fixed points are determined via
\begin{eqnarray}
     \vec w^*= (I+\xi \Omega W^*)^{-1} \frac{\vec s}{\alpha \beta}.
\end{eqnarray}

Let us assume that $\vec w=\vec w^*+\delta \vec w$, where $\vec w^*$ is a fixed point. Then, we have 
\begin{eqnarray}
    \frac{d}{dt} \delta \vec w=\partial_{\vec w} \vec f(\vec w^*) \delta \vec w.
\end{eqnarray}
For memristors one has\cite{Caravelli2018}
\begin{eqnarray}
    f_{i}(\vec w)=\alpha w_i-\sum_k (I+\xi \Omega W)^{-1}_{ik} (\Omega s)_k,
\end{eqnarray}
from which, if we use $\partial_x A^{-1}= - A^{-1} (\partial_x A) A^{-1}$
\begin{eqnarray}
    \partial_{w_j} f_i=\alpha \delta_{ij}+ \xi \frac{1}{\beta}\sum_{krts}(I+\xi \Omega W)^{-1}_{ik}\Omega_{kr} (\partial_{w_j} W)_{rt}   (I+\xi \Omega W)^{-1}_{ts}  (\Omega s)_s.
\end{eqnarray}
Evaluating this at the fixed point, we have
\begin{eqnarray}
    \alpha \beta \vec W=(I+\xi \Omega W)^{-1} \Omega\vec s
\end{eqnarray}
from which
\begin{eqnarray}
   J_{ij}= \partial_{w_j} f_i&=&\alpha \Big(\delta_{ij}+ \xi \sum_{k}(I+\xi \Omega W)^{-1}_{ik}\Omega_{kj} W_j\Big) \nonumber \\
   &=&\alpha \Big(\delta_{ij}+ \xi (I+\xi \Omega W \Omega)^{-1}_{ij} W_j\Big)
\end{eqnarray}
where the last line can be derived from the Neumann representation of the inverse and the projection condition. We now aim to prove that $J_{ij}\succ 0$ which, as $\alpha>0$ and $\xi>0$, will follow from $(I+\xi \Omega W \Omega)^{-1}_{ij} W_j\succ 0$.

Now we have that for any matrix $A$, $A\sim P A P^{-1}$, from which we obtain $AD \sim \sqrt{D} A \sqrt{D}$ for $D\succ 0$. Thus, $(I+\xi \Omega W \Omega)^{-1}_{ij} W_j\sim \sqrt{W_i}(I+\xi \Omega W \Omega)^{-1}_{ij} \sqrt{W_j} $. This matrix is clearly positive as it is symmetric and $(I+\xi \Omega W \Omega)^{-1}_{ij}$ is positive because $\Omega W \Omega$ is positive. This implies that  $J_{ij}\succ 0$ and any fixed point of the equation will be unstable.
Of course, there might a possibility that one might start from an initial condition which is a fixed point of the dynamics. 

Let us thus discuss how difficult it is to initialize the system on the fixed point manifold. 
Let $\Sigma$ be the manifold of the fixed points. Then, the probability that with a random initial condition will be the ratio of the cardinalities of the two sets, the fixed point manifold and $\mathcal C(\Sigma)$ and $\mathcal C([0,1]^N)$.
We thus ask ourselves what is $\mathcal C(\Sigma)$.
We can write the fixed point equation without loss of generality as
\begin{eqnarray}
    \vec w+\xi \Omega {\vec w}^2= \frac{\vec  s}{\alpha \beta}\equiv \vec b,
\end{eqnarray}
where $({\vec w}^2)_i=w_i^2$. The equation above can be written as a set of $N$ constraints of the form
\begin{eqnarray}
    w_i+ \xi \Omega_{ii} w_i^2- b_i + \xi \sum_{j\neq i} w_j^2=0
\end{eqnarray}
which defines the set of $N$ intersecting quadrics. The intersection of these quadrics defines an algebraic variety of degree 2. According to B\'{e}zout theorem\cite{Fulton}, for a system of well behaved polynomial equations ($N$ equations with $N$ variables) of degrees $d$ we have at most $d^N$ solutions, which is exactly $2^N$ in our case. 
However $2^N$ discrete points are a set of measure zero in $[0,1]^N$.

Naturally, this implies that if one initializes the memristors at a random initial condition in $w_i(0)\in [0,1]$, the system is very unlikely to initialize on the fixed point manifold, and thus via the unstable dynamics it must reach the boundary of the convex set $[0,1]^N$, e.g. $\{0,1\}^N$. 

\section{Analysis and comparisons}
In this section we provide evidence of the capability of memristors to significantly lower the energy as measured by the Lyapunov function.

While we have demonstrated a particular form of optimization problem that is `native' to circuits of memristors, it is common across analog systems that
 embedding an arbitrary problem into this form is difficult.. For this reason we focus on problem instances that are directly embeddable in  memristor circuits; \emph{i.e.} that arise from different circuit structures.

\subsection{The instances}

To generate instances native to memristor circuits, we formalize the optimization algorithm as a map from a circuit graph $G$ to a projection operator $\Omega(G)$.  This becomes the coupling matrix of our objective function.

\begin{figure}
    \centering
    \includegraphics[scale=1.5]{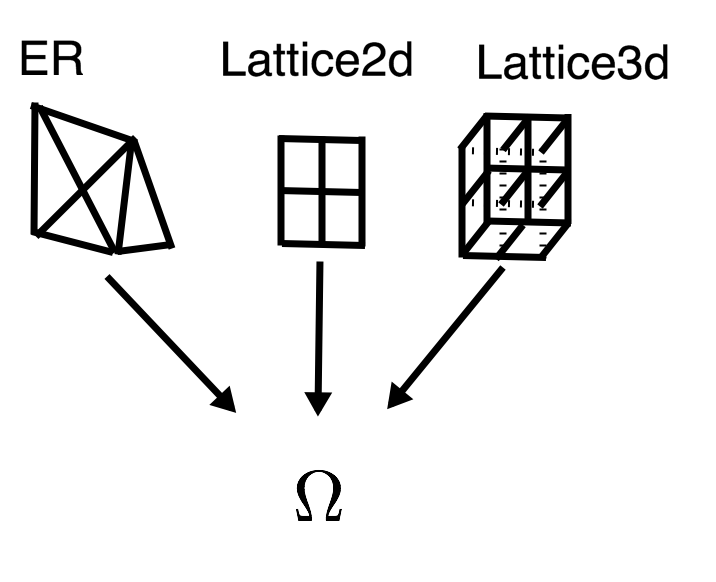}
    \caption{The three circuit instances we consider. We have an Erdos-Renyi underlying circuit (left), a 2-dimensional lattice (center)  and a 3-dimensional lattice (right). Given these, we then build the cycle matrix of the circuit $A$ and calculate the projection operator $\Omega=A^t(AA^t)^{-1} A$, which is a dense matrix, and enters in the Lyapunov function of eqn. (\ref{eq:newlyap}).}
    \label{fig:instances}
\end{figure}

The underlying graphs $G$ we chose are an Erdos-Renyi random graph (\textit{ER}), a 2-dimensional lattice (\textit{Lattice2d}) and a 3 dimensional lattice (\textit{Lattice3d}). Given these graphs, we then obtain the projection operator $\Omega_{ij}(G)=A^t(AA^t)^{-1} A$ (which is a dense matrix), which is based on the cycle space of the graph.\cite{zegarac} A graphical representation of the underlying circuit is shown in Fig. \ref{fig:instances}.

\subsection{Minimization of the continuous Lyapunov function}

\begin{figure}
    \centering
    \includegraphics[scale=0.5]{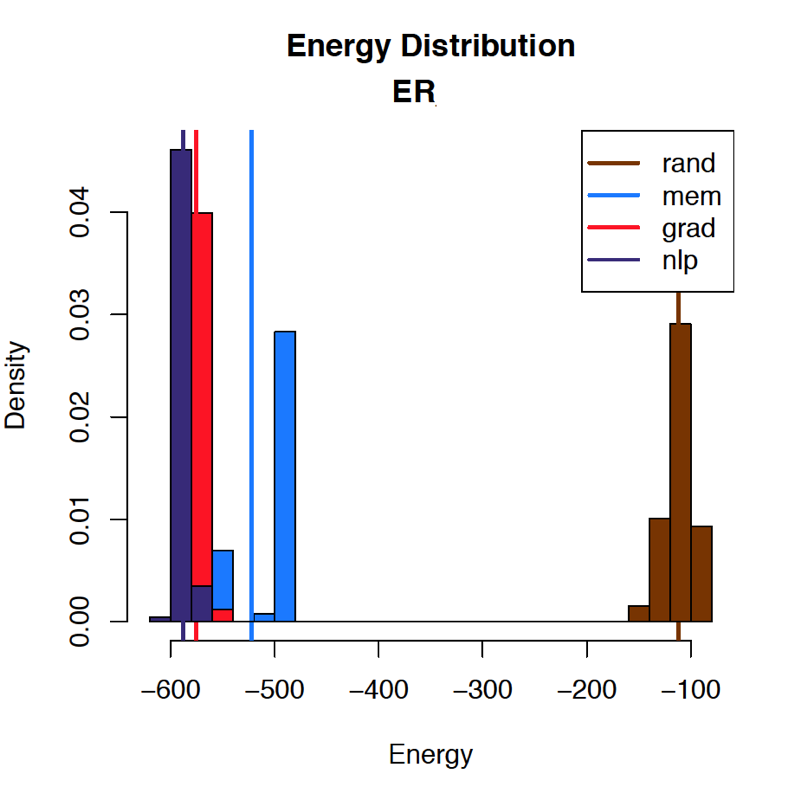}\\
    \caption{Distribution of the minima obtained with random sampling (rand), memristors (mem), gradient descent (grad) and Ipopt (nlp) for the Erdos-Renyi class (Instance 1). We see that the distribution of minima for memristors are rather close to the NLP and Grad results.}
    \label{fig:er}
\end{figure}

\begin{figure}
    \centering
    \includegraphics[scale=0.5]{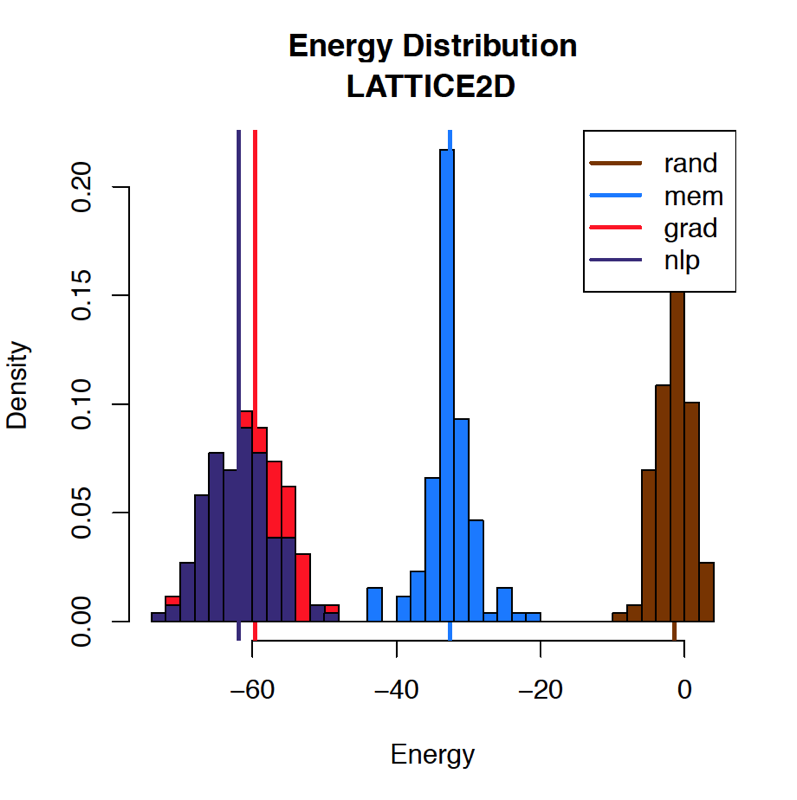}\\
    \caption{Distribution of the minima obtained with random sampling (rand), memristors (mem), gradient descent (grad) and Ipopt (nlp) for the \textit{Lattice2d} class (Instance 1). We see that the system absolute minimum is close to the tail of the non-linear programming optimization code (Ipopt), while on average these are half way between the random and nlp results.}
    \label{fig:lattice2d}
\end{figure}

\begin{figure}
    \centering
       \includegraphics[scale=0.5]{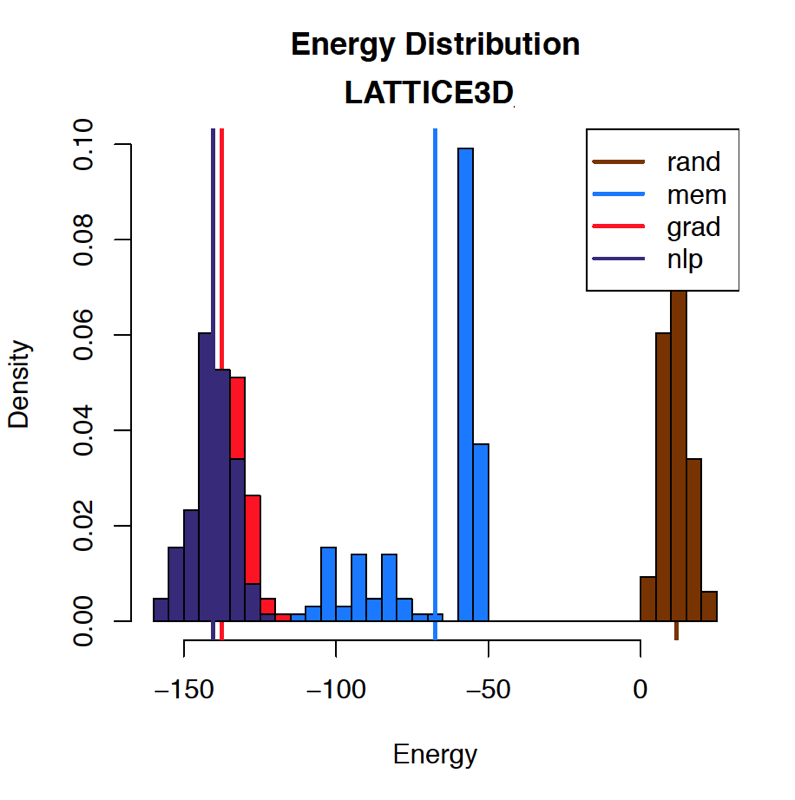}
    \caption{Distribution of the minima obtained with random sampling (rand), memristors (mem), gradient descent (grad) and Ipopt (nlp) for the \textit{Lattice3d} class (Instance 1). We see that the system absolute minimum is close to the tail of the non-linear programming optimization code (\textit{Ipopt}). }
    \label{fig:lattice3d}
\end{figure}

We compare the result of the minimization of the function
\begin{equation}
   L = -\frac{\alpha}{3} \vec{w}^T W \vec{w} - \frac{\alpha\xi}{4} \vec{w}^T W\Omega W \vec{w}
    +\frac{1}{2\beta} \vec{w}^T W\vec{x}.
    \label{eq:newlyap}
\end{equation}
using memristive circuits to other optimization algorithms. Specifically, we compare the memristive algorithm in which the dynamical equation (\ref{eq:diffeq}) is evolved numerically until it reaches a steady state, to an interior point nonlinear optimization algorithm. As a solver, we use the \textit{Ipopt} \cite{Ipopt}, an open source (second order) software for large-scale nonlinear optimization. Specifically, the software is state of the art for nonlinear problems of the form
\begin{eqnarray}
  & & \text{min}_{w \in \mathbb{R}^d} f(w)\\
  & & \text{s.t.}\ \ \ \ g^L\leq g(w) \leq g^U\\
  & & w^L\leq w\leq w^U.
\end{eqnarray}
where $f(w)$ is the function of interest (in our case equation \eqref{eq:newlyap}), $w^L$ and $w^U$ are 0 and 1 respectively in this work, and where we introduce no $g(w)$ function constraints in the optimization.
The results between the two algorithms for 15 specific instances are shown in Table 1, for the case of the ER circuits and lattices of 2- and 3- dimensions. The number of variables we consider is fairly large, e.g. in the range $N\in[112,300]$. Recognizing that both of these algorithms are sensitive to their initial starting conditions, for comparison, we consider 128 i.i.d. executions of each algorithm starting from random initial conditions in the interval $[0,1]^N$ and measure the distributions of runtime and solution quality.  In the interest of breadth, a first order optimization algorithm based on gradient descent, and a random assignment algorithm, e.g. we generate random values between $[0,1]^N$, are also included in the comparison. The results are shown in Fig. \ref{fig:er}, \ref{fig:lattice2d} and \ref{fig:lattice3d}, which compare optimization via memristor networks (\textit{mem}, light blue), random assignment (\textit{rand}, brown), Ipopt (\textit{nlp}, purple) and gradient decent (\textit{grad}, red).

First and foremost,  we note that overall \textit{Ipopt} yields the best solution quality among the optimization algorithms we considered for each specific instance. In Fig. \ref{fig:er}, \ref{fig:lattice2d} and \ref{fig:lattice3d} we plot examples of distribution of energy states for the \textit{ER, Lattice2d} and \textit{Lattice3d} cases. We see that gradient descent and Ipopt are typically close to each other for these cases, and in particular in the ER case the memristive optimization is also close to the best known solutions. For comparison, we plot in all these cases the results of a naive random optimization, from which it can be observed that the memristive circuit results are always way below the random assignment. For each class of problems we generated 5 instances. The minimum energy and average time per execution for each instance and class are shown in Tab. \ref{tab:res}. We see that memristors have a runtime advantage in terms of \textit{Ipopt} (a factor of 100), as these run much faster and one can initialize the system many times more in an equal amount of time. Also, we observe that the density of the $\Omega$ matrix places a significant computational burden for computing derivatives in second-order methods, such as \textit{Ipopt}, which is a problem feature that the memristor-based approach avoids. 

The optimal solutions found by $\textit{Ipopt}$ for the Lyapunov function confirm that (within a tolerance of $10^{-3})$ the solutions are to be found near the boundary of $[0,1]^N$.
These results somewhat confirm that the asymptotic states of a memristive circuit are to be found in local minima of a Lyapunov function, and in the discrete of the system.

\section{Conclusions}

In the present paper we have discussed the properties of memristive circuits from an optimization perspective. In particular, we have derived a new Lyapunov function for a memristive circuit of an arbitrary topology, and shown that if each memory parameter is constrained between $[0,1]$, then the asymptotic memristor values lie on the boundary of this set. This is because the fixed point of the dynamics are unstable and of countable cardinality, as we have shown. The Lyapunov function has been derived under the assumption the dynamics lies in the bulk, thus ignoring the boundaries. We have first discussed these features in the case of a single memristor device analytically.

These results have a variety of implications. \textit{In primis}, this shows that it is possible to overcome some of the problems of previously proposed Lyapunov functions in the literature. Moreover, we have tested (from the standpoint of optimization) whether analog circuits of memristors can be used for minimizing non-linear functions. We have tested three indicative classes of circuits, and found that while in none of these cases memristor dynamics obtain better minima than state of the art software (\textit{Ipopt}), they are nonetheless able to obtain good quality minima when the system is initialized multiple times. From this point of view, memristive dynamics has the advantage of providing fast good quality solutions. For instance, in the case of non-planar circuits \textit{Ipopt} took two order of magnitude longer than memristive circuits to provide a plausible minimum. In this sense, we have confirmed that memristive dynamics is naturally associated to the minimization of a Lyapunov function.  When run in hardware, we expect this speed advantage to be substantially increased.

Some comments about the difficulty of the instances we considered are in order. The class we consider, drawn from circuit structures, is previously unexplored and thus the difficulty of optimization problems in this class is unknown. We can however draw a few inferences about this class from our results. We note first that the solvers we test produce a range of potential solutions, giving evidence that these instances are not simply convex and contain a range of local minima. The software \textit{Ipopt} takes considerable time to find minima in the case of \textit{ER} and \textit{Lattice3d}, but not for \textit{Lattice2d}, in which the underlying circuit is planar. We emphasize that, while the circuit is planar, the matrix $\Omega$ is dense (none of the elements are zero). This said, it has been proven that for the case of planar circuits the matrix $\Omega$ has exponentially small support on the underlying graph \cite{Caravelli2017b}.  From this point of view, our results suggest that the class \textit{Lattice2d} is not as hard as the other two we consider due to such planarity hidden in the matrix $\Omega$; this can also be seen from the fact that \textit{Ipopt} takes significantly less time in finding good quality minima for this class.  

 While we proposed a Lyapunov function for a continuous set of variables, it is still an open question whether there exist an efficient embedding of a QUBO functional in a memristive circuit such that the QUBO functional is minimized along the dynamics as well. The key issue is that while memristors reach the boundaries of the space $\mathcal M=[0,1]^N$, it is unknown if an efficient embedding exists. This is left for future investigations.

\begin{landscape}
\vspace{3cm}
\begin{table}[b!]
\begin{tabular}{lrrrrrrrrrr}
\footnotesize
\textbf{instance} & \textbf{n} & \textbf{e} & \textbf{nlp-en} & \textbf{grad-en} & \textbf{mem-en} & \textbf{rand-en} & \textbf{nlp-tm} & \textbf{grad-tm} & \textbf{mem-tm} & \textbf{rand-tm} \\
ER\_1                  & 249   & 62001 & -601.07  & -592.09  & -577.36 & -153.90   & 118.98  & 0.29   & 1.37  &  $10^{-3}$   \\
ER\_2                 & 242   & 58564 & -576.72 & -572.52   & -561.44 & -151.70  & 103.21  & 0.21    & 2.73   &  $10^{-3}$    \\
ER\_3                 & 272   & 73441 & -671.67 & -662.14  & -654.09 & -173.22  & 159.37  & 0.29   & 3.10   &  $10^{-3}$   \\
ER\_4                  & 220   & 48400 & -520.03 & -511.88  & -498.89 & -127.31  & 76.98  & 0.20    & 1.71  &  $10^{-3}$   \\
ER\_5                  & 262   & 68644 & -650.62 & -640.00  & -630.79 & -168.76  & 139.56  & 0.29    & 1.10  &  $10^{-3}$   \\
Latt2d\_1 & 112   & 12544 & -73.80 & -70.98  & -43.82 & -8.86 & 3.96   & 0.04    & 0.48  &  $10^{-3}$   \\
Latt2d\_2 & 112   & 12544 & -78.72 & -71.99  & -41.50 & -9.42 & 3.99  & 0.04    & 0.44  &  $10^{-3}$   \\
Latt2d\_3 & 112   & 12544 & -71.86 & -68.00  & -44.69 & -3.03   & 4.10  & 0.06     & 0.68   &  $10^{-3}$   \\
Latt2d\_4 & 112   & 12544 & -73.6 & -72.33  & -42.98 & -9.30  & 4.12  & 0.04   & 0.87  &  $10^{-3}$    \\
Latt2d\_5 & 112   & 12544 & -71.99 & -70.17  & -45.62 & -8.63  & 3.99  & 0.06    & 0.75  &  $10^{-3}$   \\
Latt3d\_1 & 300   & 90000 & -158.71 & -159.01   & -110.36 & 0.00  & 164.26  & 0.40   & 1.83   &  $10^{-3}$   \\
Latt3d\_2 & 300   & 90000 & -169.60 & -164.01  & -132.64 & -9.11  & 150.42  & 0.38   & 2.52    &  $10^{-3}$    \\
Latt3d\_3 & 300   & 90000 & -161.88 & -162.91  & -113.74 & -1.59  & 140.47   & 0.41    & 4.21  & 2$\cdot 10^{-3}$   \\
Latt3d\_4 & 300   & 90000 & -164.15 & -167.83  & -116.01 & -6.74  & 145.19  & 0.31  & 3.63  &  $10^{-3}$   \\
Latt3d\_5 & 300   & 90000 & -169.36 & -166.25  & -115.71 & -9.98  & 151.80  & 0.30   & 3.33   &  $10^{-3}$  
\end{tabular}
\captionof{table}{Result of the optimization of each instance (1-5) of of the three classes considered in this article, the Erdos-Renyi (ER-random), and the \textit{Lattice2d} and \textit{Lattice3d} Classes. The number of nodes of the circuit are given in \textit{nodes} column (n), while the number of edges of the graph (the variables) in the \textit{edge} column (e). The results of the optimization using \textit{Ipopt}, gradient descent, memristors and random are in the columns (\textit{nlp,grad,mem,rand})-en  columns respectively, while the average time (in seconds) for the solution to be obtained in the (\textit{nlp,grad,mem,rand})-tm  columns.} 
\label{tab:res}
\end{table}
\end{landscape}

\section{Acknowledgments}
This work was carried out under the auspices of the NNSA of the U.S. DoE at LANL under Contract No. DE-AC52-06NA25396, in particular via DOE-ER grants PRD20190195. Also, FCS is supported by a CNLS Fellowship.

\bibliographystyle{ws-rv-van}
\bibliography{ws-rv-sample}


\end{document}